# Resolution-enhanced parallel coded ptychography for high-throughput optical imaging

Shaowei Jiang[1, 4], Chengfei Guo[1, 4*], Pengming Song[1], Niyun Zhou[2], Zichao Bian[1], Jiakai Zhu[1], Ruihai Wang[1], Pei Dong[2], Zibang Zhang[1], Jun Liao[2], Jianhua Yao[2], Bin Feng[1], Michael Murphy[3], and Guoan Zheng[1*]

[1]Department of Biomedical Engineering, University of Connecticut, Storrs, CT 06269, USA
[2]Tencent AI lab, Shenzhen, Guangdong 02215, China
[3]Immunopathology Laboratory, Department of Dermatology, University of Connecticut Health Center, Farmington, CT 06030, USA
[4]These authors contributed equally to this work.

*Correspondence:
C. G. (chengfei.guo@uconn.edu) or G. Z. (guoan.zheng@uconn.edu)

**Abstract:** Ptychography is an enabling coherent diffraction imaging technique for both fundamental and applied sciences. Its applications in optical microscopy, however, fall short for its low imaging throughput and limited resolution. Here, we report a resolution-enhanced parallel coded ptychography technique achieving the highest numerical aperture and an imaging throughput orders of magnitude greater than previous demonstrations. In this platform, we translate the samples across the disorder-engineered surfaces for lensless diffraction data acquisition. The engineered surface consists of chemically etched micron-level phase scatters and printed sub-wavelength intensity absorbers. It is designed to unlock an optical space with spatial extent $(x, y)$ and frequency content $(k_x, k_y)$ that is inaccessible using conventional lens-based optics. To achieve the best resolution performance, we also report a new coherent diffraction imaging model by considering both the spatial and angular responses of the pixel readouts. Our low-cost prototype can directly resolve 308-nm linewidth on the resolution target without aperture synthesizing. Gigapixel high-resolution microscopic images with a 240-mm$^2$ effective field of view can be acquired in 15 seconds. For demonstrations, we recover slow-varying 3D phase objects with many 2π wraps, including optical prism and convex lens. The low-frequency phase contents of these objects are challenging to obtain using other existing lensless techniques. For digital pathology applications, we perform accurate virtual staining by using the recovered phase as attention guidance in a deep neural network. Parallel optical processing using the reported technique enables novel optical instruments with inherent quantitative nature and metrological versatility.

**Introduction**

When a crystalline specimen is illuminated with a homogeneous coherent field, the reciprocal space will be populated with periodic Bragg peaks. The intensity of the Bragg peaks can be recorded using light detector. The phase associated with each peak, however, is lost in the acquisition process, preventing the reconstruction of the real-space structure through Fourier synthesis. The original concept of ptychography was developed to address the phase problem of crystallography in electron microscopy[1]. By translating a narrow coherent probe beam on the specimen, it aspires to extract the phase of Bragg peaks from the beating patterns at the reciprocal space. In 2004, the iterative phase retrieval framework[2] was adopted for ptychographic reconstruction, thereby bringing the technique to its modern form[3-13]. The experiment procedure remains the same: the specimen is laterally translated through a spatially confined probe and the Fourier diffraction patterns are recorded at the reciprocal space. Different from the original concept, the reconstruction process iteratively imposes two different sets of constraints. The diffraction measurements serve as the Fourier magnitude constraints in the reciprocal space. The confined probe beam limits the physical extent of the object for each measurement and serves as the support constraint in the real space. Since ptychography does not require a reference beam and can image extended samples without compact object support, it has rapidly attracted interest from the coherent diffraction imaging (CDI) community. In the past decade, it has become an indispensable imaging tool in most synchrotron and national laboratories worldwide[14].

The applications of ptychography in optical microscopy, however, fall short for its low imaging throughput and limited resolution. Under either the confined probe[15-17] or the full-field illumination condition[10, 18, 19], the highest numerical aperture (NA) of lensless ptychography demonstrated is no better than 0.4. The imaging throughput is orders of magnitude lower than that of the automated microscope platform such as the whole slide scanner. Current implementations also require frequent correction of the complex probe beam due to stability issues of the coherent light source and other system perturbations. The blind recovery of both the complex object and probe beam becomes a challenging task if the complex object contains slow-varying phase contents. Therefore, aside from a few impressive proof-of-principle demonstrations, ptychographic CDI remains of limited practical application in the visible region.

It is possible to combine ptychography with a lens for optical microscopy[9, 20, 21]. Fourier ptychography is one example that swaps the real and reciprocal space using a lens[9]. It employs an LED array to illuminate the sample from different incident angles. At each angle, a different part of the object's Fourier spectrum passes through the pupil aperture for detection. The use of angle-varied illumination, however, requires the object to be a thin slice[22]. Recovering objects with slow-varying phase profiles is a challenging task as the low-frequency phase information cannot be effectively converted into intensity variations for detection. This same challenge also applies to other common lensless CDI techniques, including blind ptychography with both object and probe unknown[3-5, 23], support-constraint approach[24-26], multi-height and multi-wavelength approaches[26-29], digital in-line holography[26, 30, 31], and transport-of-intensity approach[32].

Reaping the benefits of ptychography in the visible region would require the above challenges to be properly addressed. Here we report a resolution-enhanced parallel coded ptychography technique achieving the highest numerical aperture of ~0.8 and an imaging throughput orders of magnitude higher than previous demonstrations. In our platform, we replace the objective lens with a disorder-engineered surface that serves as an unconfined computational scattering lens[33-40]. The engineered surface consists of chemically etched micron-level phase scatters and printed sub-wavelength intensity absorbers. It is designed to unlock an optical space with spatial extent $(x, y)$ and spatial frequency content $(k_x, k_y)$ that is inaccessible using conventional lens-based optics. Object exit waves with large diffraction angles can be converted into smaller angles by the engineered surface, thereby facilitating super-resolution imaging beyond the limit imposed by the system transfer function. We permanently attach the surface to the image sensor, with the engineered facet facing the pixels. As such, the surface is protected from direct contact with the object, addressing the intrinsic instability issue related to system perturbations. Once characterized, the device can be operated without the need to update the surface profile, avoiding the blind ptychography problem and more importantly, addressing the challenge of recovering slow-varying phase profiles.

To achieve the highest resolution performance, we also develop a new CDI model to consider both the spatial and angular responses of the pixel readouts. In this model, the incoherent nature of the signal integration process is characterized by the spatial response of the image sensor. The coherent nature of the angle-sensitive detection process, on the other hand, is characterized by the angular response of the pixels. With the engineered surface and the new CDI model, we directly resolve the 308-nm linewidth on the resolution target using a 1.85-µm pixel size detector, obtaining the highest NA among all ptychographic demonstrations. Using the prototype platform, we acquire gigapixel microscopic images with a ~240-mm$^2$ effective FoV in 15 seconds, demonstrating an imaging throughput orders of magnitude higher than those in previous implementations. For comparison, a state-of-the-art whole slide scanner can acquire microscopic images with a 225-mm$^2$ FoV and sub-micron resolution in ~2 mins. Rapid and precise autofocusing, however, remains a challenge during the sample scanning process[41].

The imaging performance of our device is validated by a wide range of applications. For metrology applications, we recover the quantitative height and phase of different 3D targets, including the slow-varying phase profiles that are challenging to obtain using other existing lensless imaging techniques. For digital pathology applications, we develop a phase-attention-guided deep neural network for virtual staining. We perform rapid whole slide imaging and locate the virtually stained white blood cells (WBCs) over a centimeter FoV. The rich structural information and superior phase sensitivity in ptychographic reconstruction also allow us to perform accurate fluorescence labeling after training on the same type of specimen. The reported technique provides a turnkey and scalable solution for optical microscopy with inherent quantitative nature and metrological versatility. Parallel optical processing using a coded ptychographic array also enables novel instruments for high-throughput label-free imaging.

**Results**
**Coded ptychography via disorder-engineered surface.** Figure 1a shows the operating principle of the coded ptychography technique, where the sample is translated across the disorder-engineered surface for diffraction data acquisition. The engineered surface consists of two types of structures in Figure 1b: micron-level phase scatters and sub-wavelength absorbers. The phase scatters are created by chemically etching the glass surface and they enable fast convergence of the recovery process (Figures S1-S2). The small intensity absorbers are created by printing the carbon nanoparticles on the etched surface, and they can effectively redirect large-angle diffractions into smaller angles for detection. In this technique, the engineered surface encodes the otherwise inaccessible object information into intensity variations for detection. This operation is similar to that of the scattering lens[33-37], where the high-frequency object information is down modulated into the passband of the optical system for detection. Object translation above the engineered surface allows the exit waves to be encoded by different parts of the surface with translation diversity[3], justifying the proposed name of coded ptychography.

The design of the disorder-engineered surface also shares its roots with the metasurface used for wide-field imaging[39]. Both approaches use 2D thin surface as scattering lens to provide optical 'randomness' of conventional disordered media but in a way that is fully known a priori. The metasurface approach uses a 2D array of subwavelength phase scatters that are designed to have a large angular correlation range. The subwavelength phase scatters, however, have stability issues for the recovery process, especially when the complex object has a relatively large phase range (Figures S1-S2). In contrast, micron-level phase scatters enable fast and stable first-loop convergence. In our design, we integrate the micron-level phase scatters with small intensity absorbers on the same surface, achieving the best performance as summarized in Figure S2. In Figure S3, we also show that a lower transmittance of the absorbers enables a better modulation process for phase retrieval. The use of carbon nanoparticles is well aligned with this observation. Another key difference between our engineered surface and the metasurface[39] is the fabrication process. Metasurface with sub-wavelength phase scatters often requires the use of electron beam lithography for defining the small patterns on photoresist. In contrast, we employ a lithography-free fabrication process to make the disorder-engineered surface, enabling turnkey and scalable manufacturing at low cost (Figure S4). Figure S5 shows the captured images of the surface using an oil-immersion lens.

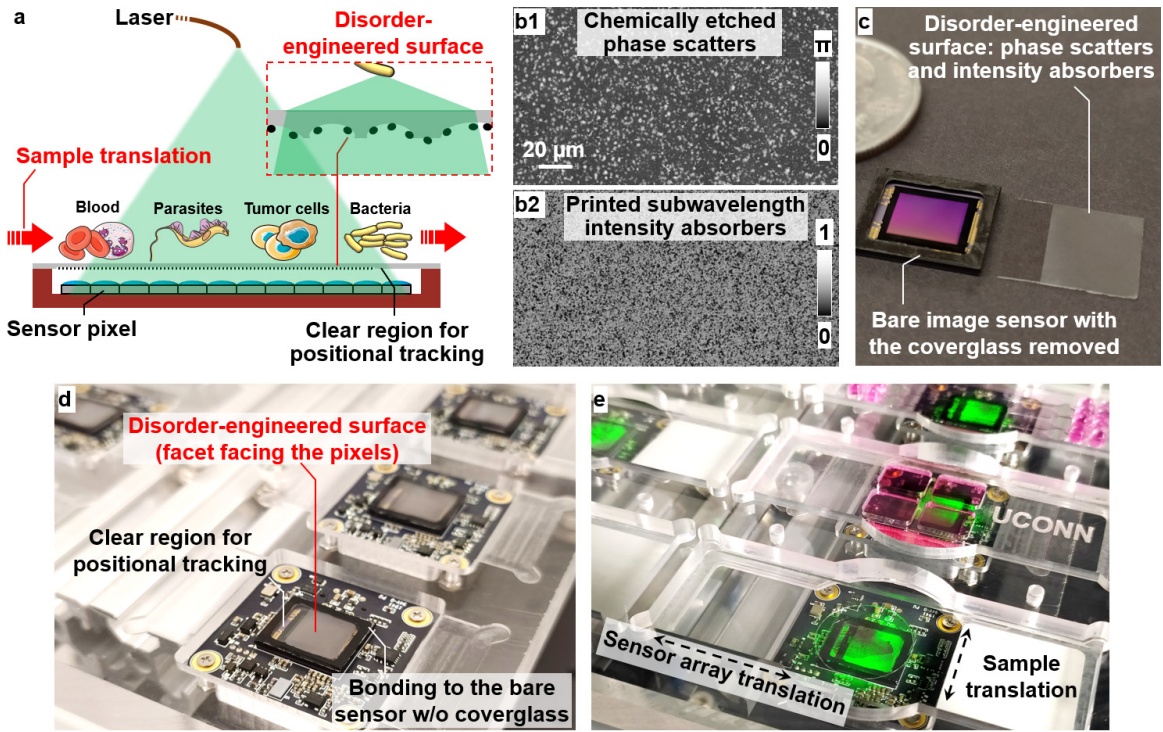

**Figure 1.** Parallel coded ptychography for high-throughput lensless imaging. (a) A disorder-engineered surface is attached to an image sensor for encoding the otherwise inaccessible object information for detection. The samples are translated across the engineered surfaces for diffraction data acquisition. Inset shows the engineered surface that contains chemically-etched micron-level phase scatters and printed subwavelength intensity absorbers. It is designed to unlock an optical space with spatial extent ($x$, $y$) and frequency content ($k_x$, $k_y$) that is inaccessible using conventional lens-based optics. (b) The recovered phase and intensity maps of the disorder-engineered surface. (c) The original coverglass of the image sensor is removed and replaced by the engineered surface. (d) The integrated device with the engineered facet facing the sensor pixels, addressing the stability issue of disordered media and enabling long-term operation without recalibration. A transparent region on the engineered surface is used for sample positional tracking. (e) The reported device has the flexibility to image different types of samples. Green light shows the laser beams impinging on the samples for large FoV lensless diffraction data acquisition (~30 mm$^2$ FoV for each sensor).

Figure 1c shows the bare image sensor where we remove its original coverglass and replace it with the engineered surface. Figure 1d shows the integrated device with the engineered facet faces the pixels. This arrangement avoids direct contact between external objects and the surface, addressing the intrinsic instability issue related to system perturbations. In comparison, stability of conventional disordered media is only several hours[18, 37, 38]; realignment of both the position and orientation is often needed for each experiment[18, 37-39]. Once characterized, the reported device can be operated without recalibrating the surface profile, avoiding the blind ptychography problem where both the object and probe need to be jointly recovered[3-5, 23]. Figure 1e shows the flexibility of using the reported device for imaging different types of samples.

The turnkey fabrication of the device allows the implementation in parallel. Figure 2a shows the design of the parallel coded ptychographic array using 8 sensors and Figure 2b shows the prototype platform. The assembling procedures can be found in Supporting Information, Video 1, and Note 1. To operate the system, we translate the sample (or sensor) for diffraction data acquisition. The step size in-between adjacent acquisitions is 1-3 µm, corresponding to a translation speed of ~60 µm/s. The sample can be in continuous motion without motion blur. Time gated acquisition or pulse illumination in automated microscopy is not needed in our approach. For one FoV of the image sensor (~30 mm$^2$), we typically acquire 200-450 images in 8-15 s. We then position the sample to a new FoV and repeat the acquisition process (Supporting Information, Video 2). 8 sensors operating in parallel enable the acquisition of a 240-mm$^2$ effective FoV in less than 15 s.

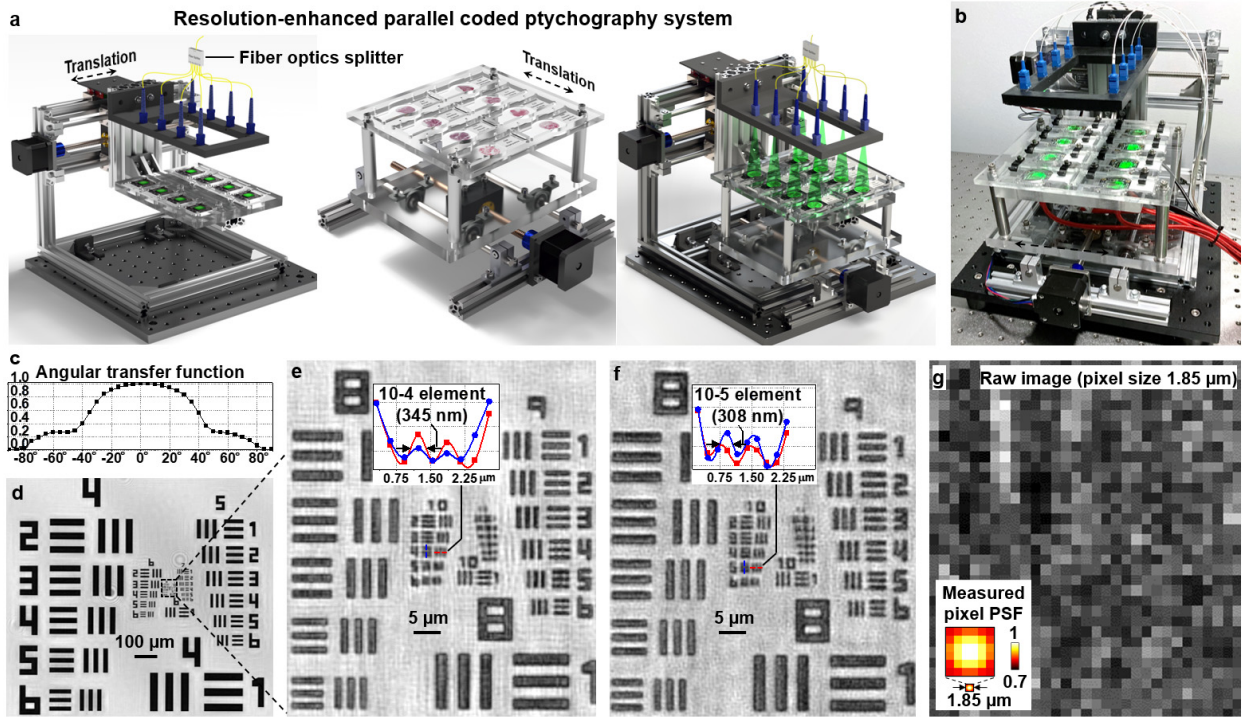

**Figure 2.** Parallel coded ptychography prototype and its resolution-enhanced imaging performance. (a) Design of the coded ptychographic array where a 1-to-8 fiber beam splitter is used to couple laser light for sample illumination. Refer to Supporting Information, Video 1, and Note 1 for its operation. (b) The prototype with 8 sensors operating in parallel, achieving the highest throughput for wide-field, high-resolution microscopy. (c) Pixel readout at different incident angles. (d-e) The recovered image of a resolution target using 532 nm incident light, resolving the 345-nm linewidth. (f) The recovered image using 405 nm incident light, resolving the 308-nm linewidth. (g) The captured raw image for g, with a pixel size of 1.85 µm. Inset shows the measured intra-pixel intensity response (also refer to Figure S6).

**Imaging model with spatial and angular responses.** Conventional CDI approach models the pixel output as an integration of the wave intensity across its active sensing area. This model, however, fails to consider the coherent nature of the impinging light waves. It is only valid when the pixel angular response is uniform across the spatial frequency bandwidth of the impinging waves. To achieve a high-NA performance, we need to consider both the incoherent intensity integration and the coherent angular response in the CDI model as follows:

$$I_j(x - x_j, y) = Avg_{M \times M}\{|[W(x, y - y_j) \cdot S(x - x_j, y) \otimes PSF_{free}(d)] \otimes PSF_{angular}|^2 \cdot Mask_{pixel}(x - x_j, y)\}, \quad (1)$$

where $I_j$ is the j$^{th}$ captured image, $W$ is the object exit wave on the engineered surface, $S$ is the transmission profile of the engineered surface, $PSF_{free}(d)$ is the point spread function (PSF) for free-space propagation of distance $d$ (~320 μm in our device), $PSF_{angular}$ is a complex PSF for modeling the coherent angular response of the pixels and its Fourier transform is the angular transfer function shown in Figure 2c, $Mask_{pixel}$ is the incoherent spatial intensity response mask of the pixels, '·' represent point-wise multiplication, '⊗' represents convolution, $x_j$ is the positional shift of the sensor array, $y_j$ is the positional shift of the sample array, and '$Avg_{M \times M}$' represents $M \times M$ average pooling for image down-sampling. For Figure 2c, we mount the image sensor on a rotation stage and illuminate the sensor with a plane wave. The intensity response of the pixels is then plotted as the function of the rotation angle. The concept of $Mask_{pixel}$ is also demonstrated in Figure S6, where it models both the intra-pixel response (i.e., pixel PSF) and the inter-pixel response across the entire FoV (i.e., fixed pattern noise and non-uniformity gain of different pixels). In the calibration process, we measure the $Mask_{pixel}$ using the slanted-edge approach[42] (Methods and Figure S18).

The precise estimation of positional shifts $x_j$ and $y_j$ is the key to achieving the highest resolution in ptychographic imaging. We use the images captured through the coded clear region of the engineered surface for estimating the initial positional shifts[43] and further refine them in the iterative process (Supporting Information, Note 2). With the positional shifts and the premeasured $S$, $PSF_{angular}$, and $Mask_{pixel}$, we can recover the object exit wave $W$ on the engineered surface and then digitally propagate it back to any plane along the axial direction. Figures 2d-f show the recovered images of the resolution target under 532-nm and 405-nm laser illumination. We resolve 345-nm and 308-nm linewidth on the target and the corresponding best NA is ~0.8, the highest among all lensless ptychographic demonstrations. As a reference, Figure 2g shows a raw diffraction measurement with a raw pixel size of 1.85 μm. To the best of our knowledge, it is the first time to consider both the coherent angular response ($PSF_{angular}$) and the incoherent spatial response ($Mask_{pixel}$) for high-resolution CDI. In comparison, Figure S7 shows the recovered images assuming a conventional CDI model with uniform angular and spatial responses. The best-achieved resolution has been degraded to 435-nm and 390-nm respectively, ~25% worse than our results in Figure 2e-f.

**Imaging slow-varying 3D phase objects with many 2π wraps.** Imaging slow-varying 3D phase objects with many 2π wraps is a challenging task for conventional lensless imaging techniques, where the slow-varying phase information cannot be effectively converted into intensity variations for detection. For example, in lensless in-line setups[26-32], if we place an optical prism on top of an image sensor, the captured intensity will be uniform across the entire image and contains no information of the phase. Thus, it is impossible to restore the linear phase ramp in the phase retrieval process. One key advantage of the reported technique is the permanent attachment of the engineered surface to the sensor. With the pre-measured surface, the low-spatial-frequency object phase can be effectively converted into spatial distortions in the diffraction patterns. Thus, the device enables true quantitative phase recovery regardless of the object's spatial frequency contents, enabling large-scale phase and height measurement for various metrology applications.

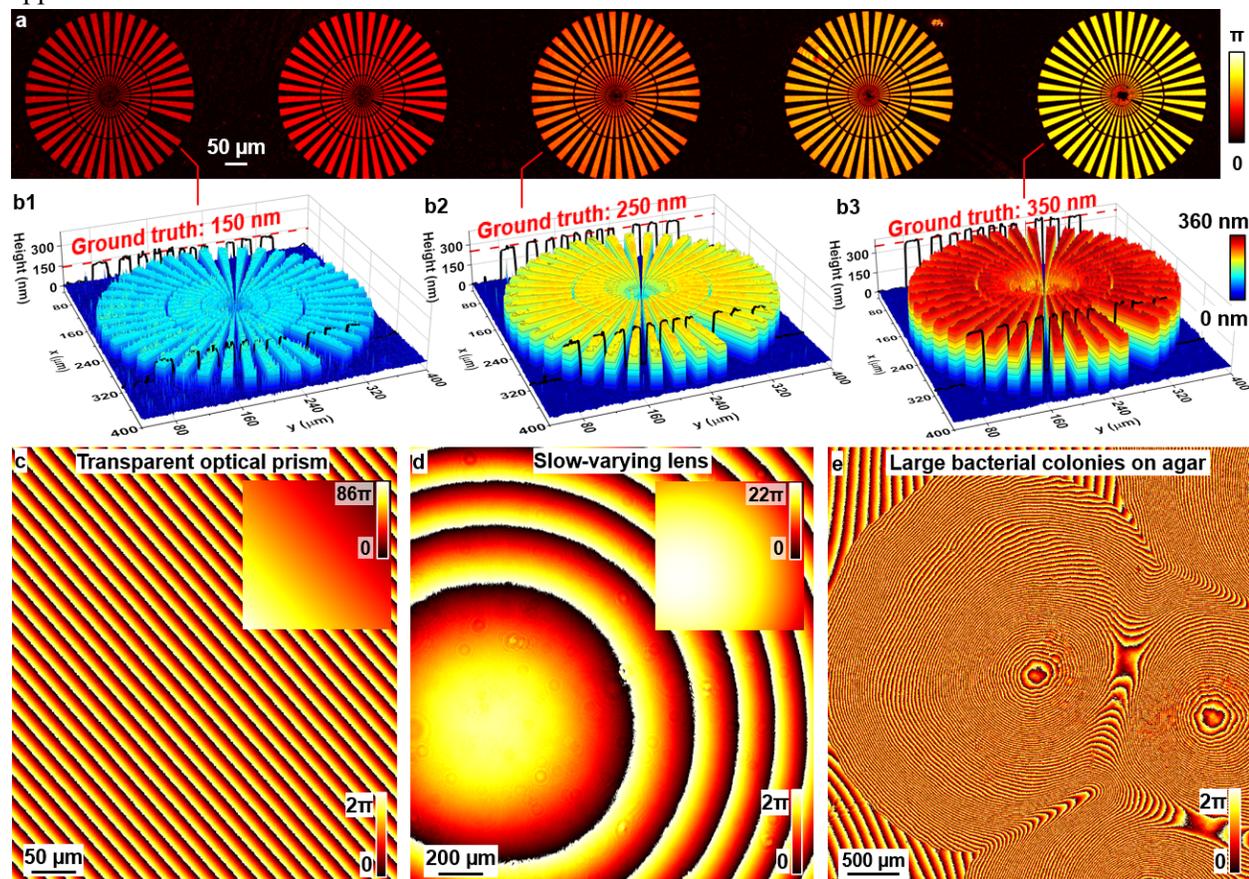

**Figure 3.** Quantitative phase and height metrology. (a) Recovered phase image of a phase target. (b) The heightmap for the phase target. The recovered phase profiles of an optical prism (c), bi-convex lens (d), and live bacterial colonies on an uneven agar plate (e). These slow-varying phase profiles with many $2\pi$ wraps are challenging to obtain using other common lensless on-chip microscopy techniques.

In Figure 3a, we show the recovered phase of a quantitative phase target. The corresponding height maps are shown in Figure 3b, where the line traces match well with the ground truths. Figures 3c-3e show the recovered phases of slow-varying phase objects, including a 2-degree optical prism, a bi-convex lens with a 5-cm focal length, and bacterial colonies on an uneven agar plate. In comparison, Figure S8 shows the failed reconstructions of the same objects with blind ptychography[4, 5], where both the object's exit wave and the engineered surface profile are jointly updated in the process. In Figure S9, we also show the recovered whole slide phase image of a histology slide at centimeter scale.

**Virtual staining for high-throughput whole slide imaging.** The reported platform can also be used for high-throughput whole slide imaging (WSI) for digital pathology, where tissue slides are converted to digital images for inspection and computer-aided diagnosis. To facilitate our platform for WSI, we develop an unsupervised deep neural network, termed phase-attention-guided cycle-consistent generative adversarial network (pcGAN), for virtual color and fluorescence staining (Figures 4 and S10). Our pcGAN employs the cycleGAN architecture[44] and uses the recovered phase image for attention guidance in the image translation process (Methods). As shown in Figure 4a, one generator in the pcGAN takes the recovered intensity and phase as the input and produces the virtually stained images according to different color and fluorescence staining styles.

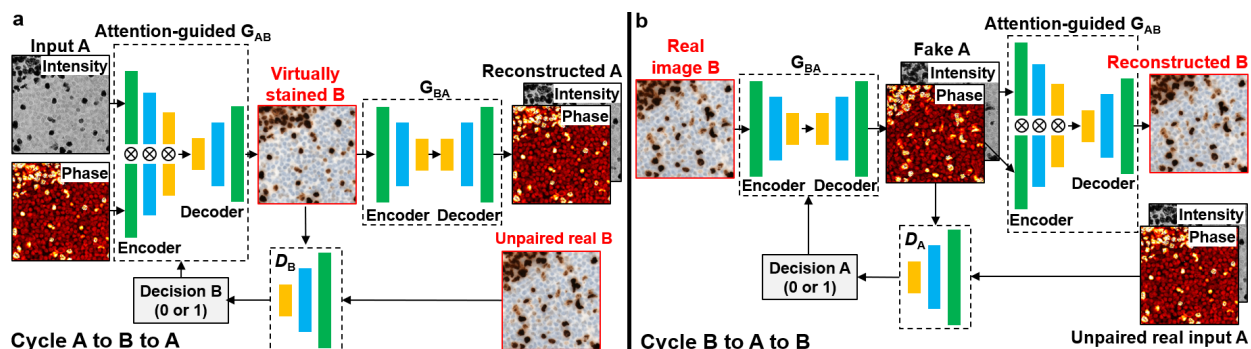

**Figure 4.** The phase-attention-guided cycle-consistent generative adversarial network (pcGAN) for virtual staining. In the generator $G_{AB}$, we use the recovered phase as the attention guidance for addressing the intensity ambiguity problem (Figure S11). (a) The A-B-A cycle structure for translating the input ptychographic reconstruction $A$ into virtually stained image $B$, then back to $A$. (b) The B-A-B cycle-structure for translating the real stained image $B$ into ptychographic reconstruction $A$, then back to $B$. Refer to Figure S10 for the detailed network structures.

There are three innovations for the reported pcGAN. First, we use the recovered phase as the attention map for constructing the generator. This attention-guided strategy enables accurate virtual staining and avoids the ambiguity issue of intensity-only or phase-only input. Figure S11 shows the comparison of virtually stained images with and without using the phase attention map. Without using the phase for attention guidance, cells with similar intensity result in ambiguity for staining, leading to staining errors in Figure S11b. Second, our network does not require paired images for training and it is different from the previously reported supervised approach[45]. The alignment and registration of paired data are not needed in our implementation. In the training process of pcGAN, we use two sets of unpair data: the recovered intensity and phase from our platform, and the captured color image using a WSI system[46]. The pcGAN learns the mapping from the two unpaired datasets. Third, we employ a multiscale structural similarity index measure (SSIM) loss term to avoid color reversal and feature distortion between the input and output (Methods).

To recover the whole slide image using the reported platform, we need to properly refocus the exit wave to the in-focus positions. In contrast with the conventional WSI where a focus map is generated before scanning, our platform allows autofocusing after the data has been acquired. As shown in Figure 5a, we generate focus maps of the samples based on the recovered exit waves (Supporting Information, Note 2). In Figure 5b, we train 5 pcGANs to generate the virtually stained images according to different staining styles. We can resolve the kinetoplast, flagellum, and nucleus of the Trypanosoma parasites in Figure 5b2. As a reference, Figure 5c shows the ground-truth images captured using a 40×, 0.95 NA objective lens. The difference between our images and the ground truths is shown in Figure 5d and quantified by SSIM. Figures S12-S16 and Videos 3-7 show the whole slide images of those in Figure 5b. In Figure 6a, we also perform virtual Wright staining of the recovered complex blood smear image using our prototype. Figure 6b shows the zoom-in views of the WBCs and the corresponding virtually stained color images. The ground truth images are shown in Figure 6c for comparison. In Figure 6d, we show the first 45 white blood cell images automatically located via cell segmentation.

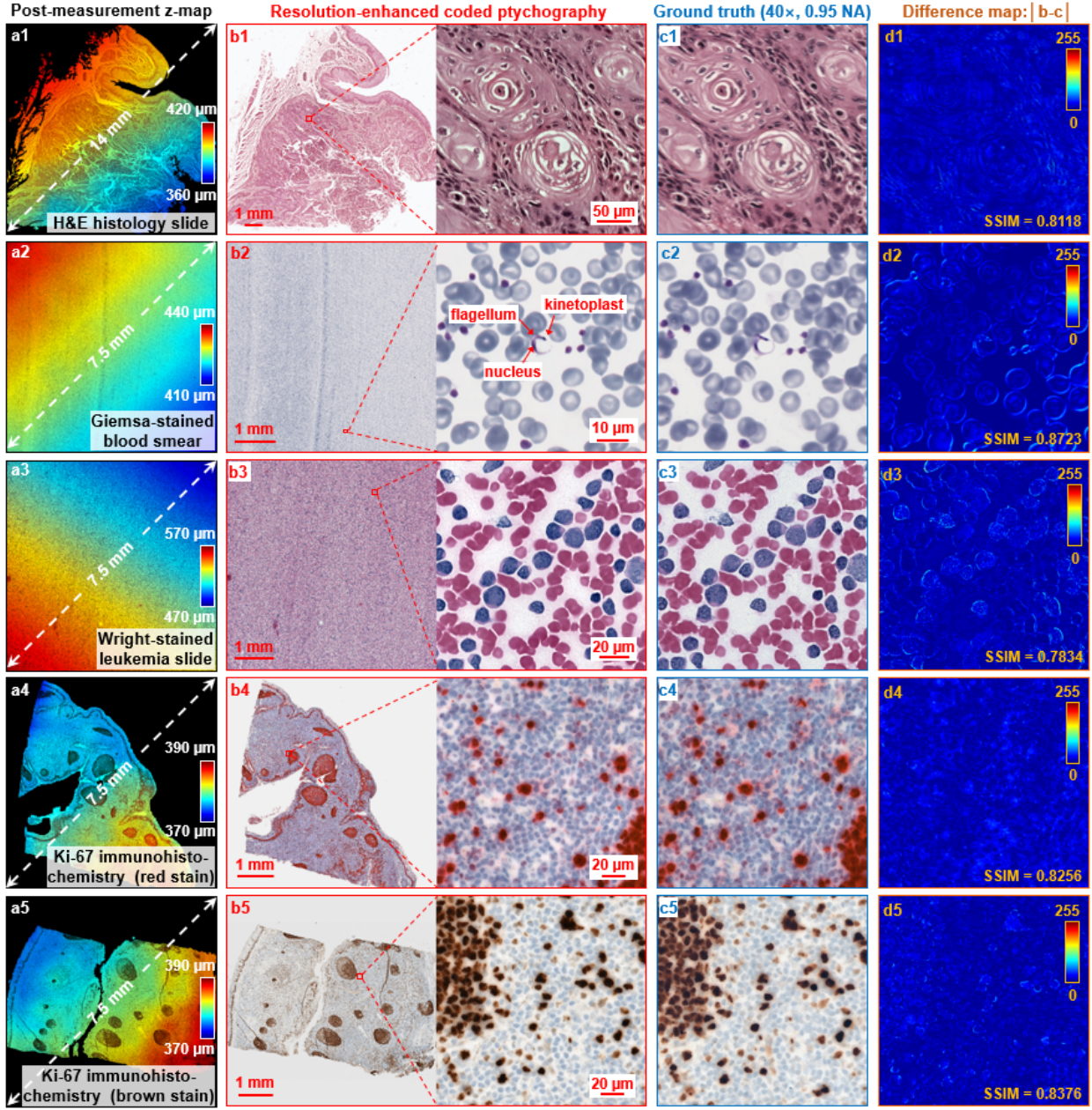

**Figure 5.** High-throughput whole slide imaging using the reported parallel coded ptychography platform. (a) The focus map for WSI, where we propagate the recovered exit wave to different axial positions and select the focus points based on a focus metric. (b) The virtually stained whole slide images using our system. (c) The ground-truth images captured by a 40×, 0.95 NA lens. (d) The difference between b and c. Refer to Figures S12-16 and Videos 3-7 for the whole slide images.

Similar to virtual color staining in Figures 5-6, we also perform virtual fluorescence staining in Figure 7, where the recovered whole slide image of mouse kidney slide is virtually stained with Alexa Fluor 488 for labeling glomeruli and convoluted tubules. Figure S17 shows the difference between the virtually stained images and the ground truth fluorescence images acquired by a regular microscope. The rich structural information and superior phase sensitivity in ptychographic reconstruction allow accurate fluorescence labeling using the pcGAN.

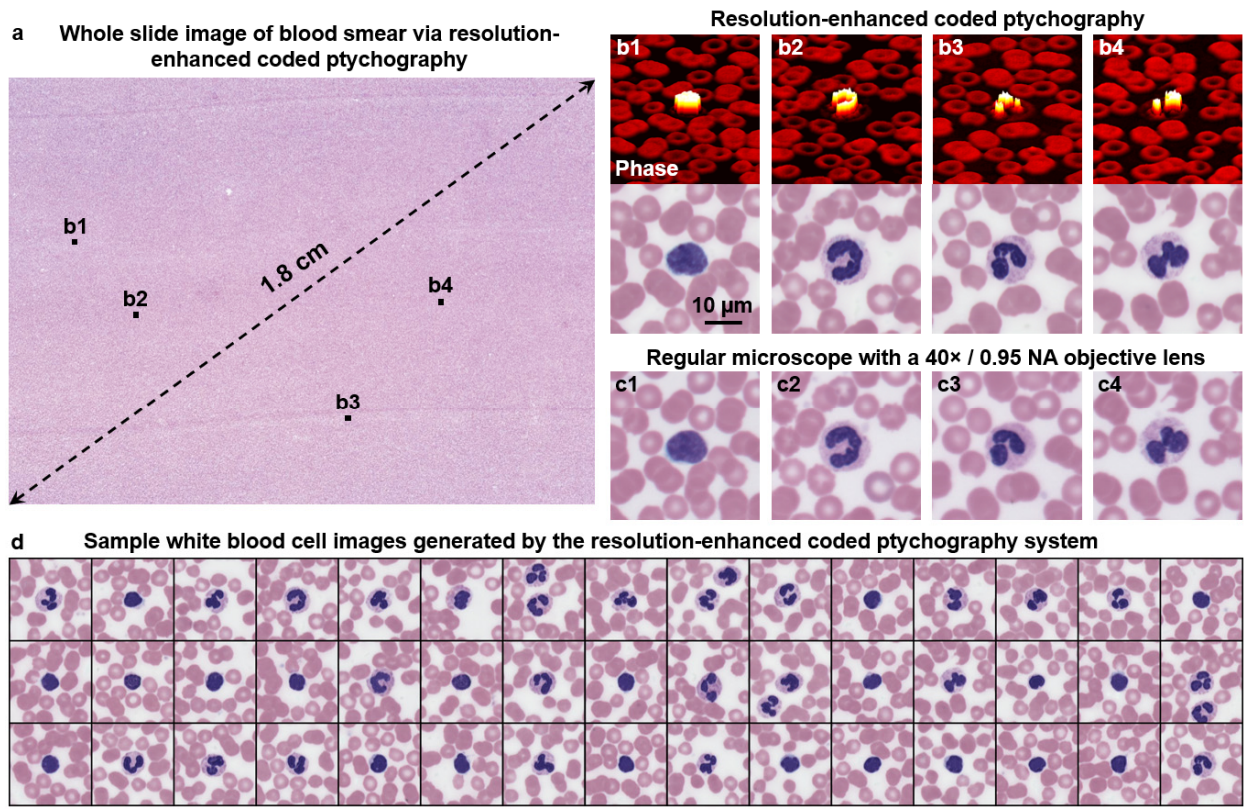

**Figure 6.** Blood film inspection based on the virtually stained image. (a) The virtually stained whole slide image of a blood smear. (b) The zoom-in view of the WBCs' phase images and the corresponding stained images. (c) The ground truth images captured using a 40×, 0.95 NA objective. (d) Sample virtually stained WBCs tracked by the platform.

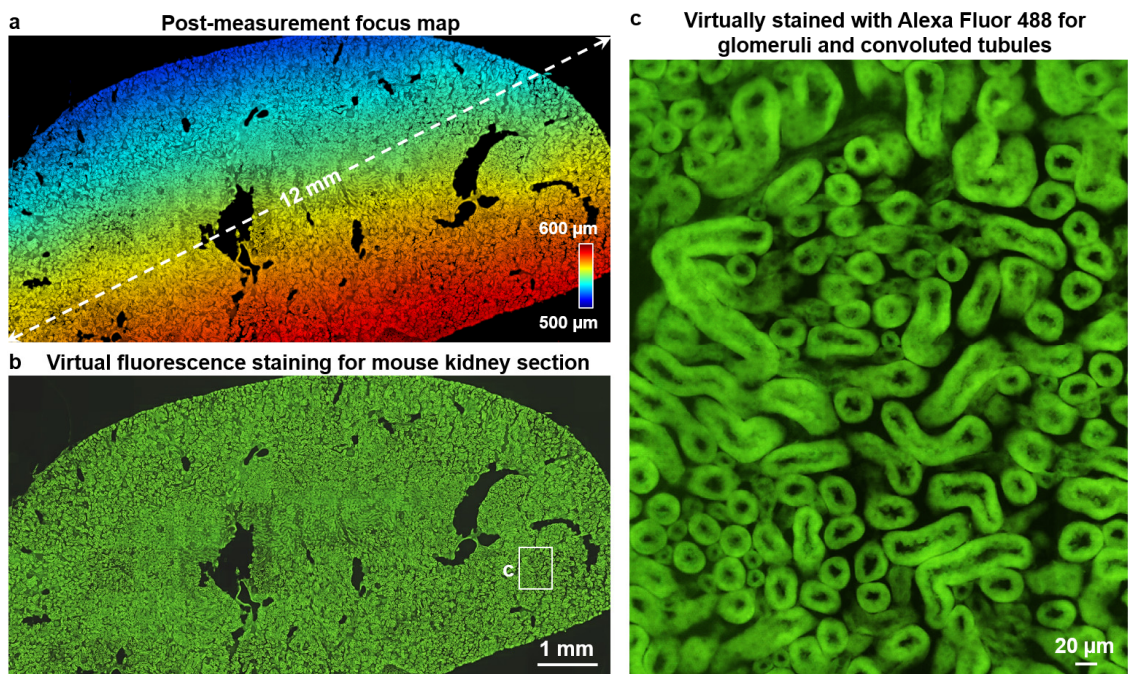

**Figure 7.** Virtual fluorescence labeling using whole slide ptychographic reconstruction. (a) The generated focus map post measurement. (b) The virtually stained whole slide image of mouse kidney section. (c) The magnified view of b, where we digitally label the glomeruli and convoluted tubules using the pcGAN.

**Discussion and conclusion**

The scaling of complexity in array microscopy is a major obstacle impeding large-scale, high-throughput imaging for biomedical applications. Here, we demonstrate a turnkey parallel coded ptychography system with an imaging throughput orders of magnitude higher than the previous implementations. The contributions and innovations of the reported platform can be summarized as follows. First, the disorder-engineered surface contains both chemically-etched phase scatters and printed intensity absorbers for achieving the highest NA among all existing ptychographic systems. The achieved NA and resolution is at least two times higher than those demonstrated in wide-field lensless ptychographic systems[18]. The unique fabrication process also differs from that of the conventional metasurface, allowing cost-effective manufacturing without involving optical or electron lithography. Second, we develop a new CDI model that considers both the incoherent intensity integration mask and the coherent angular response of the pixels. This imaging model outperforms the regular CDI model where both intensity and angular responses are assumed to be uniform. Third, the permanent attachment of the engineered surface to the image sensor avoids the stability and alignment issues of the conventional scattering lens. With the engineered facet facing the pixels, the attachment process generates an integrated device that eliminates potential damages from external contacts. It also addresses the blind ptychography problems where both the object and probe need to be updated in the process[18]. In this regard, we demonstrate the recovery of slow-varying phase profiles with many $2\pi$ wraps, including optical prism, convex lens, and large bacterial colonies. The low-frequency phase contents of these objects are challenging to obtain using existing lensless techniques[26-32]. Fourth, lens-based system often suffers from demanding requirements on the mechanical stability. Images can be easily defocused due to misalignment or environmental vibrations. In our device, we encode an empty region on the engineered surface for positional tracking. The device can be operated without feedback from the mechanical stage, enabling open-loop optical acquisition. The approach using an encoded clear region for positional tracking can also be used in other ptychographic implementations to eliminate the need for precise mechanical scanning. Post-measurement autofocusing demonstrated in this work further eliminates the need to maintain a precise distance between the sample and the engineered surface. Fifth, we employ a novel unsupervised pcGAN for virtual staining. The network uses phase image as attention guidance and the training process requires no paired data. Compared with the cycleGAN virtual staining process[47], the phase attention guidance employed in our platform addresses the ambiguity problem of intensity- or phase-only networks. Lastly, the parallel image acquisition using a coded ptychographic array allows us to achieve an imaging throughput higher than that of a regular whole slide scanner while at a small fraction of the cost.

The fundamental resolution limit of the reported platform is that corresponding to an NA of 1 in air. The current achieved NA is close to 0.8 in air. To further improve the NA, one straightforward solution is to add immersion oil in-between the object and the engineered surface. We also note that the resolution performance demonstrated here is achieved without aperture synthesizing. It is also possible to improve the resolution of 2D thin objects using the angle-varied illumination concept demonstrated in Fourier ptychography[9] and lensless in-line holography[48, 49]. However, for 3D objects such as an optical prism, tilting the illumination plane-wave would change the spectrum of the object rather than just shifting it in the Fourier space. In a wave optics picture, tilting the illumination would rock the Ewald sphere around the origin. As such, each captured image corresponds to a spherical cap of the Ewald sphere in the Fourier space[50]. These spherical caps do not span a single 2D plane in the Fourier domain. Our ongoing effort is to perform 3D quantitative diffraction tomographic reconstruction via angle-varied illumination[51].

In our design of the engineered surface, the micron-level relatively large phase scatters can better convert the object phase information into intensity variations for detection. The principle is similar to that of phase contrast microscopy, where a phase plate is used to convert the phase information into intensity variations for better visualization. The sub-wavelength intensity absorbers, on the other hand, can convert the large-angle diffractions into smaller angles for detection. The principle is similar to that of structured illumination microscopy, where a non-uniform illumination pattern is used to modulate the high-frequency object information into the passband of the optical system. We also note that the optimal choice of the coded pattern (i.e., the surface profile) may depend on the object exit waves. In the current study, we have not explored all possible combinations of different intensity absorbers and phase scatters. How to choose the optimal pattern profile and its convergence property is an important topic for the reported technique. Further research along this direction is highly desired.

Ptychography is an enabling CDI technique for both fundamental and applied sciences. Its applications in high-throughput optical imaging, currently in their early stage, will continue to improve in performance and expand in applications. The reported study is among the first steps in this direction.

**Methods**

**Preparation of the disorder-engineered surface.** The engineered surface contains micron-size phase scatters with sub-wavelength intensity absorbers. To prepare the surface, we first cut a microscope coverslip into the proper size for covering the bare image sensor (Sony IMX 226). We then applied the glass etching chemicals (17% barium sulfate, 11% sulfuric acid, 8% sodium bifluoride, 5% ammonium bifluoride) on the surface for 1-3 s before washing them with water. A longer etching time often leads to large phase scatters that cannot be modeled as a thin 2D surface. This etching and cleaning process was repeated 5-10 times for generating the dense phase scatters on the surface. The average depth and size of the phase scatters are 320 nm and 3.6 µm in Figure 1b, and the same parameters are used in Figure S1-S2 for validation. In the etching process, part of the surface was protected by a plastic film for positional tracking. With the etched surface, we printed carbon particles as intensity absorbers on the etched facet as shown in Figure S4. We first rubbed the etched surface with silk clothes to make it positively charged. We then used a negatively charged printer roller to deposit carbon nanoparticles on the etched surface. The resulting engineered surface was permanently attached to the bare image sensor using nail polish, a procedure inspired by the coverslip sealing of microscope slide (Figure S4c-S4e). With the integrated devices, we mounted the sensor array and the sample holder on two separated motorized stages modified from a low-cost computer numerical control (CNC) router[46].

**Measuring the pixel PSF.** In Eq. (1), we used $Mask_{pixel}$ to model the incoherent intensity response of the pixels. For each pixel of the captured raw image, we have $M$ by $M$ subpixels in the $Mask_{pixel}$ as the pixel PSF for super-resolution imaging. The intensity response of the pixel PSF is higher at the center than that at the edge (Figure S6). In our implementation, we measured the pixel PSF using the slanted-edge approach[42], where a slightly tilted edge of an opaque target was projected on the image sensor using a 20×, 0.75 NA objective. Figure S18 shows the measurement process and the result of the measured pixel PSF.

**The recovery process.** The recovery process is based on the imaging model of Eq. (1), where each raw pixel in the captured image corresponds to $M \times M$ pixels in the exit wave $W$. In the calibration experiment, we used an H&E slide as a weak-phase object that does not contain slow-varying phase features. We acquired 1500 raw images for joint recovery of the object exit wave $W(x, y)$ and the surface profile $S(x, y)$. For all subsequent experiments, the calibrated surface profile $S(x, y)$ and the measured $Mask_{pixel}(x, y)$ were enforced in the phase retrieval process for avoiding the blind ptychography problem. The number of acquisitions ranges from 200 to 450 and the images were captured at 30 frames per second. The setup of the calibration experiment is the same as that of the non-blind one; the only difference is that we acquire more images for the calibration experiment. For all experiments, we also iteratively refine the positional shift $x_j$, $y_j$ and correct the laser intensity for each acquired image in the recovery process (Supporting Information, Note 2).

**Phase-attention-guided network for virtual staining.** The reported pcGAN in Figure 4 consists of two generators, $G_{AB}$ and $G_{BA}$, and two discriminators, $D_A$ and $D_B$. For the recovered amplitude and phase images $A$ using the coded ptychography system, the generator produces a virtually stained image $B$ of the object ($G_{AB}: A \rightarrow B$). Similarly, the generator $G_{BA}$ creates a fake amplitude and phase images $A$ based on a real color image $B$ captured by the whole slide scanner ($G_{BA}: B \rightarrow A$). Each generator has a corresponding discriminator, which attempts to tell apart the generated images from the real ones, i.e., $D_A$ distinguishes $A$ from $G_{BA}(B)$, and $D_B$ distinguishes $B$ from $G_{AB}(A)$. We use the following loss function in our training process:

$Loss(G_{AB}, G_{BA}, D_A, D_B) = L_{GAN-AB}(G_{AB}, D_B, A, B) + L_{GAN-BA}(G_{BA}, D_A, B, A) + \lambda_1 \cdot L_{cyc}(G_{AB}, G_{BA}) + \lambda_2 \cdot (1 - msSSIM_g(G_{AB}(A), A)) + \lambda_2 \cdot (1 - msSSIM_g(G_{BA}(B), B))$,

where $L_{GAN-AB}$ and $L_{GAN-BA}$ are the adversarial losses for the forward and backward generator-discriminator pairs, and $L_{cyc}$ is the cycle consistency loss. The term $msSSIM_g$ represents the multiscale SSIM loss between the green channel of the stained images and the intensity of the ptychographic reconstruction. We introduce this loss to avoid color reversal and distortion between the input and output images[47]. $\lambda_1$ and $\lambda_2$ are the weights for different loss terms and we set them to 10 and 0.2, respectively. The network structure of the generator $G_{AB}$ is shown in Figure S10a, which consists of 9 pairs of down-sampling blocks followed by 9 up-sampling blocks. The down-sampling blocks in the phase path are used as multiscale attention guidance for the feature maps in the intensity path. The red arrows in Figure S10a indicate skip connection and concatenation of the feature maps from the down-sampling blocks to the up-sampling blocks. For the generator $G_{BA}$ in Figure S10b, we use a U-net structure that consists of 9 down-sampling blocks followed by 9 up-sampling blocks. For the discriminators $D_A$ and $D_B$, we employ the PatchGAN classifiers in Figure S10c.

**Acknowledgment**
G. Z. acknowledges the support of the National Science Foundation 1510077, 1700941, 2012140, and the UConn SPARK grant. P. S. acknowledges the support of the Thermo Fisher Scientific fellowship.

**Author contributions**
S. J. developed the prototype system and prepared the display items. C. G. fabricated the disorder-engineered surface with phase scatters and intensity absorbers. P. S. and Z. B. measured the angular response. J. Z. and R. W. validated the initial concept and tested the engineered surface. N. Z., P. D., J. L., and S. J. developed the phase-attention-guided neural network for unsupervised virtual staining. G. Z. conceived the idea and supervised the project. All authors contributed to manuscript preparation and editing.

**Notes**
The authors declare no competing financial interest.

**Supporting Information**
The Supporting Information is available free of charge at Simulations of the disorder-engineered surface, fabrication process, reconstructions with uniform spatial and angular responses, comparison with blind ptychography, coded ptychographic whole slide imaging, the network structure of pcGAN, network performance with and without phase attention, recovered gigapixel images of biospecimens, virtual fluorescence labeling, pixel point spread function measurement, assembling procedures of the coded ptychography platform, reconstruction process.

**Reference:**
(1) Hoppe, W.; Strube, G., Diffraction in inhomogeneous primary wave fields. 2. Optical experiments for phase determination of lattice interferences. *Acta Crystallogr. A* **1969,** *25*, 502-507.
(2) Faulkner, H. M. L.; Rodenburg, J., Movable aperture lensless transmission microscopy: a novel phase retrieval algorithm. *Physical review letters* **2004,** *93* (2), 023903.
(3) Guizar-Sicairos, M.; Fienup, J. R., Phase retrieval with transverse translation diversity: a nonlinear optimization approach. *Optics express* **2008,** *16* (10), 7264-7278.
(4) Thibault, P.; Dierolf, M.; Bunk, O.; Menzel, A.; Pfeiffer, F., Probe retrieval in ptychographic coherent diffractive imaging. *Ultramicroscopy* **2009,** *109* (4), 338-343.
(5) Maiden, A. M.; Rodenburg, J. M., An improved ptychographical phase retrieval algorithm for diffractive imaging. *Ultramicroscopy* **2009,** *109* (10), 1256-1262.
(6) Dierolf, M.; Menzel, A.; Thibault, P.; Schneider, P.; Kewish, C. M.; Wepf, R.; Bunk, O.; Pfeiffer, F., Ptychographic X-ray computed tomography at the nanoscale. *Nature* **2010,** *467* (7314), 436-439.
(7) Maiden, A. M.; Humphry, M. J.; Zhang, F.; Rodenburg, J. M., Superresolution imaging via ptychography. *JOSA A* **2011,** *28* (4), 604-612.
(8) Thibault, P.; Menzel, A., Reconstructing state mixtures from diffraction measurements. *Nature* **2013,** *494* (7435), 68-71.
(9) Zheng, G.; Horstmeyer, R.; Yang, C., Wide-field, high-resolution Fourier ptychographic microscopy. *Nature photonics* **2013,** *7* (9), 739.
(10) Stockmar, M.; Cloetens, P.; Zanette, I.; Enders, B.; Dierolf, M.; Pfeiffer, F.; Thibault, P., Near-field ptychography: phase retrieval for inline holography using a structured illumination. *Scientific reports* **2013,** *3* (1), 1-6.
(11) Jiang, Y.; Chen, Z.; Han, Y.; Deb, P.; Gao, H.; Xie, S.; Purohit, P.; Tate, M. W.; Park, J.; Gruner, S. M., Electron ptychography of 2D materials to deep sub-ångström resolution. *Nature* **2018,** *559* (7714), 343-349.
(12) Gardner, D. F.; Tanksalvala, M.; Shanblatt, E. R.; Zhang, X.; Galloway, B. R.; Porter, C. L.; Karl Jr, R.; Bevis, C.; Adams, D. E.; Kapteyn, H. C., Subwavelength coherent imaging of periodic samples using a 13.5 nm tabletop high-harmonic light source. *Nature Photonics* **2017,** *11* (4), 259-263.
(13) Hruszkewycz, S.; Allain, M.; Holt, M.; Murray, C.; Holt, J.; Fuoss, P.; Chamard, V., High-resolution three-dimensional structural microscopy by single-angle Bragg ptychography. *Nature materials* **2017,** *16* (2), 244-251.
(14) Pfeiffer, F., X-ray ptychography. *Nature Photonics* **2018,** *12* (1), 9-17.
(15) Maiden, A. M.; Rodenburg, J. M.; Humphry, M. J., Optical ptychography: a practical implementation with useful resolution. *Optics letters* **2010,** *35* (15), 2585-2587.
(16) Balaur, E.; Cadenazzi, G. A.; Anthony, N.; Spurling, A.; Hanssen, E.; Orian, J.; Nugent, K. A.; Parker, B. S.; Abbey, B., Plasmon-induced enhancement of ptychographic phase microscopy via sub-surface nanoaperture arrays. *Nature Photonics* **2021,** *15* (3), 222-229.
(17) Li, P.; Maiden, A., Lensless LED matrix ptychographic microscope: problems and solutions. *Applied optics* **2018,** *57* (8), 1800-1806.
(18) Jiang, S.; Zhu, J.; Song, P.; Guo, C.; Bian, Z.; Wang, R.; Huang, Y.; Wang, S.; Zhang, H.; Zheng, G., Wide-field, high-resolution lensless on-chip microscopy via near-field blind ptychographic modulation. *Lab on a Chip* **2020,** *20* (6), 1058-1065.

(19) Song, P.; Wang, R.; Zhu, J.; Wang, T.; Bian, Z.; Zhang, Z.; Hoshino, K.; Murphy, M.; Jiang, S.; Guo, C., Super-resolved multispectral lensless microscopy via angle-tilted, wavelength-multiplexed ptychographic modulation. *Optics Letters* **2020,** *45* (13), 3486-3489.
(20) Marrison, J.; Räty, L.; Marriott, P.; O'toole, P., Ptychography–a label free, high-contrast imaging technique for live cells using quantitative phase information. *Scientific reports* **2013,** *3* (1), 1-7.
(21) McDermott, S.; Maiden, A., Near-field ptychographic microscope for quantitative phase imaging. *Optics express* **2018,** *26* (19), 25471-25480.
(22) Dong, S.; Horstmeyer, R.; Shiradkar, R.; Guo, K.; Ou, X.; Bian, Z.; Xin, H.; Zheng, G., Aperture-scanning Fourier ptychography for 3D refocusing and super-resolution macroscopic imaging. *Optics express* **2014,** *22* (11), 13586-13599.
(23) Ou, X.; Zheng, G.; Yang, C., Embedded pupil function recovery for Fourier ptychographic microscopy. *Optics express* **2014,** *22* (5), 4960-4972.
(24) Fienup, J. R., Reconstruction of a complex-valued object from the modulus of its Fourier transform using a support constraint. *JOSA A* **1987,** *4* (1), 118-123.
(25) Miao, J.; Charalambous, P.; Kirz, J.; Sayre, D., Extending the methodology of X-ray crystallography to allow imaging of micrometre-sized non-crystalline specimens. *Nature* **1999,** *400* (6742), 342-344.
(26) Latychevskaia, T., Iterative phase retrieval for digital holography: tutorial. *JOSA A* **2019,** *36* (12), D31-D40.
(27) Greenbaum, A.; Ozcan, A., Maskless imaging of dense samples using pixel super-resolution based multi-height lensfree on-chip microscopy. *Optics express* **2012,** *20* (3), 3129-3143.
(28) Greenbaum, A.; Zhang, Y.; Feizi, A.; Chung, P.-L.; Luo, W.; Kandukuri, S. R.; Ozcan, A., Wide-field computational imaging of pathology slides using lens-free on-chip microscopy. *Science translational medicine* **2014,** *6* (267), 267ra175-267ra175.
(29) Bao, P.; Zhang, F.; Pedrini, G.; Osten, W., Phase retrieval using multiple illumination wavelengths. *Optics letters* **2008,** *33* (4), 309-311.
(30) Xu, W.; Jericho, M.; Meinertzhagen, I.; Kreuzer, H., Digital in-line holography for biological applications. *Proceedings of the National Academy of Sciences* **2001,** *98* (20), 11301-11305.
(31) Mudanyali, O.; Tseng, D.; Oh, C.; Isikman, S. O.; Sencan, I.; Bishara, W.; Oztoprak, C.; Seo, S.; Khademhosseini, B.; Ozcan, A., Compact, light-weight and cost-effective microscope based on lensless incoherent holography for telemedicine applications. *Lab on a Chip* **2010,** *10* (11), 1417-1428.
(32) Gureyev, T. E.; Nugent, K. A., Rapid quantitative phase imaging using the transport of intensity equation. *Optics communications* **1997,** *133* (1-6), 339-346.
(33) Vellekoop, I. M.; Lagendijk, A.; Mosk, A., Exploiting disorder for perfect focusing. *Nature photonics* **2010,** *4* (5), 320-322.
(34) Choi, Y.; Yang, T. D.; Fang-Yen, C.; Kang, P.; Lee, K. J.; Dasari, R. R.; Feld, M. S.; Choi, W., Overcoming the Diffraction Limit Using Multiple Light Scattering in a Highly Disordered Medium. *Physical Review Letters* **2011,** *107* (2), 023902.
(35) van Putten, E. G.; Akbulut, D.; Bertolotti, J.; Vos, W. L.; Lagendijk, A.; Mosk, A., Scattering lens resolves sub-100 nm structures with visible light. *Physical review letters* **2011,** *106* (19), 193905.
(36) Park, J.-H.; Park, C.; Yu, H.; Park, J.; Han, S.; Shin, J.; Ko, S. H.; Nam, K. T.; Cho, Y.-H.; Park, Y., Subwavelength light focusing using random nanoparticles. *Nature photonics* **2013,** *7* (6), 454.
(37) Choi, Y.; Yoon, C.; Kim, M.; Choi, W.; Choi, W., Optical imaging with the use of a scattering lens. *IEEE Journal of Selected Topics in Quantum Electronics* **2013,** *20* (2), 61-73.
(38) Park, J.; Park, J.-H.; Yu, H.; Park, Y., Focusing through turbid media by polarization modulation. *Optics letters* **2015,** *40* (8), 1667-1670.
(39) Jang, M.; Horie, Y.; Shibukawa, A.; Brake, J.; Liu, Y.; Kamali, S. M.; Arbabi, A.; Ruan, H.; Faraon, A.; Yang, C., Wavefront shaping with disorder-engineered metasurfaces. *Nature photonics* **2018,** *12* (2), 84-90.
(40) Lee, K.; Park, Y., Exploiting the speckle-correlation scattering matrix for a compact reference-free holographic image sensor. *Nature communications* **2016,** *7* (1), 1-7.

(41) Bian, Z.; Guo, C.; Jiang, S.; Zhu, J.; Wang, R.; Song, P.; Zhang, Z.; Hoshino, K.; Zheng, G., Autofocusing technologies for whole slide imaging and automated microscopy. *Journal of Biophotonics* **2020,** *13* (12), e202000227.
(42) Estribeau, M.; Magnan, P. In *Fast MTF measurement of CMOS imagers using ISO 12333 slanted-edge methodology*, Detectors and Associated Signal Processing, International Society for Optics and Photonics: 2004; pp 243-252.
(43) Bian, Z.; Jiang, S.; Song, P.; Zhang, H.; Hoveida, P.; Hoshino, K.; Zheng, G., Ptychographic modulation engine: a low-cost DIY microscope add-on for coherent super-resolution imaging. *Journal of Physics D: Applied Physics* **2019,** *53* (1), 014005.
(44) Zhu, J.-Y.; Park, T.; Isola, P.; Efros, A. A. In *Unpaired image-to-image translation using cycle-consistent adversarial networks*, Proceedings of the IEEE international conference on computer vision, 2017; pp 2223-2232.
(45) Rivenson, Y.; Liu, T.; Wei, Z.; Zhang, Y.; de Haan, K.; Ozcan, A., PhaseStain: the digital staining of label-free quantitative phase microscopy images using deep learning. *Light: Science & Applications* **2019,** *8* (1), 1-11.
(46) Guo, C.; Bian, Z.; Jiang, S.; Murphy, M.; Zhu, J.; Wang, R.; Song, P.; Shao, X.; Zhang, Y.; Zheng, G., OpenWSI: a low-cost, high-throughput whole slide imaging system via single-frame autofocusing and open-source hardware. *Optics Letters* **2020,** *45* (1), 260-263.
(47) Wang, R.; Song, P.; Jiang, S.; Yan, C.; Zhu, J.; Guo, C.; Bian, Z.; Wang, T.; Zheng, G., Virtual brightfield and fluorescence staining for Fourier ptychography via unsupervised deep learning. *Optics Letters* **2020,** *45* (19), 5405-5408.
(48) Luo, W.; Greenbaum, A.; Zhang, Y.; Ozcan, A., Synthetic aperture-based on-chip microscopy. *Light: Science & Applications* **2015,** *4* (3), e261-e261.
(49) Luo, W.; Zhang, Y.; Feizi, A.; Göröcs, Z.; Ozcan, A., Pixel super-resolution using wavelength scanning. *Light: Science & Applications* **2016,** *5* (4), e16060-e16060.
(50) Zheng, G.; Shen, C.; Jiang, S.; Song, P.; Yang, C., Concept, implementations and applications of Fourier ptychography. *Nature Reviews Physics* **2021,** *3* (3), 207-223.
(51) Horstmeyer, R.; Chung, J.; Ou, X.; Zheng, G.; Yang, C., Diffraction tomography with Fourier ptychography. *Optica* **2016,** *3* (8), 827-835.

For Table of Contents Use Only

# Resolution-enhanced parallel coded ptychography for high-throughput optical imaging

Shaowei Jiang[1, 4], Chengfei Guo[1, 4*], Pengming Song[1], Niyun Zhou[2], Zichao Bian[1], Jiakai Zhu[1], Ruihai Wang[1], Pei Dong[2], Zibang Zhang[1], Jun Liao[2], Jianhua Yao[2], Bin Feng[1], Michael Murphy[3], and Guoan Zheng[1*]

[1]Department of Biomedical Engineering, University of Connecticut, Storrs, CT 06269, USA
[2]Tencent AI lab, Shenzhen, Guangdong 02215, China
[3]Immunopathology Laboratory, Department of Dermatology, University of Connecticut Health Center, Farmington, CT 06030, USA
[4]These authors contributed equally to this work.

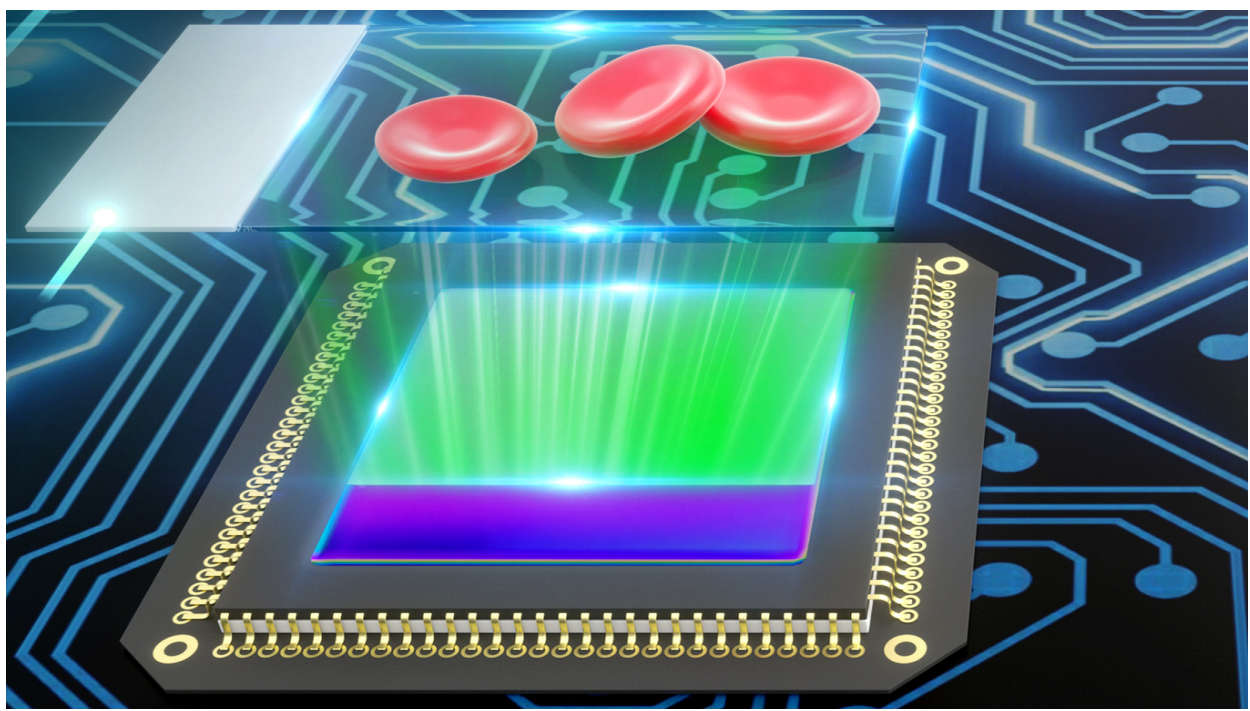

The image presents the concept of parallel coded ptychography technique. The biospecimens are translated across the disorder-engineered surfaces on top of the image sensors for diffraction data acquisition. The prototype device can resolve 308-nm linewidth on the resolution target and acquire gigapixel microscopic images with a 240-mm$^2$ effective field of view in 15 seconds. The imaging throughput is orders of magnitude higher than that of whole slider scanner while at a small fraction of the cost.